\documentclass[proceedings]{JHEP} 

\usepackage{epsfig}               

\def\beq{\begin{equation}}
\def\eeq{\end{equation}}
\def\beqa{\begin{eqnarray}}
\def\eeqa{\end{eqnarray}}
\def\a {{\rm f}}
\def\T{T}
\def\U{U} 
\def\blocA{\Gamma_{3 \times 3}}

\def\bc{\begin{center}}
\def\ec{\end{center}}

\conference{Corfu Summer Institute on Elementary Particle Physics, 1998}

\title{Soft gluon resummation for heavy quark and dijet cross sections}

\author{Nikolaos Kidonakis\thanks{Present address: Department of Physics,
Florida State University, Tallahassee, FL 32306-4350, USA.}\\
Department of Physics and Astronomy,
University of Edinburgh,
Edinburgh EH9 3JZ, Scotland, United Kingdom \\}

\abstract{We discuss the resummation of threshold logarithms for heavy
quark and dijet cross sections in hadronic collisions. 
The resummed cross sections are presented at next-to-leading logarithmic 
accuracy in terms of anomalous dimension matrices which describe the
factorization of soft gluons from the hard scattering. 
We apply our formalism to the calculation of the top quark 
production cross section at the Fermilab Tevatron.}
 
\begin{document}

\section{Introduction}

Threshold resummations have recently been studied for a variety of 
hadronic processes, in particular heavy quark and jet production
(for a review see Ref.~\cite{NK}).
Near threshold for the production of the final state 
in these processes one finds large logarithmic corrections 
which originate from soft gluon emission off the partons in the
hard scattering. These logarithms can be resummed to all orders
in perturbative QCD. 
The resummation of the leading theshold logarithms in heavy quark
and dijet production cross sections   
follows the earlier study of the Drell-Yan process \cite{St87,CT}
since these logarithms originate from the incoming partons and 
are thus universal. 
 
The resummation of next-to-leading loga-\break rithms (NLL) 
for QCD hard scattering and heavy quark production,
in particular, was presented in Refs. \cite{Thesis,KS}. 
At NLL accuracy one has to take into
account the color exchange in the hard scattering.
The resummation is formulated in terms of soft anomalous
dimension matrices which describe the factorization of soft gluons
from the hard scattering.
Applications to top and bottom quark production 
at a fixed center-of-mass scattering angle,
$\theta=90^{\circ}$ (where the relevant anomalous dimension matrices
are diagonal), 
were discussed in Ref. \cite{NKJSRV}. More recently the 
total cross section was calculated 
for top quark production at the Tevatron \cite {NKRV}.
Resummation for single-particle inclusive cross sections
has been considered in Ref.~\cite{LOS}.

The NLL resummation formalism for the \break hadroproduction of 
dijets has been presented in Refs. \cite{KOS1,KOS2}.
Jet production involves additional complications relative
to heavy quark production because of the presence of the final-state jets.

\section{Resummation for heavy quark \break production}

We begin with the resummation formalism for heavy quark production. 
First, we write the factorized form of the cross section
and identify singular distributions in it near threshold. Then we
refactorize the cross section into functions associated with gluons 
collinear to the incoming quarks,  non-collinear soft gluons,
and the hard scattering. Resummation follows from the renormalization
properties of these functions. The resummed cross section 
is written in terms of exponentials of 
anomalous dimension matrices for each partonic subprocess involved. 

\subsection{Factorized cross section}

We study the production of a pair of heavy quarks of momenta
$p_1$, $p_2$,
in collisions of hadrons $h_a$ and $h_b$ with momenta
$p_a$ and $p_b$, 
\beq
h_a(p_a)+h_b(p_b) \rightarrow {\bar Q}(p_1) +Q(p_2) + X \, ,
\eeq
with total rapidity $y$
and scattering angle $\theta$ in the pair center-of-mass frame.
The heavy quark production cross section can be written in a 
factorized form as a convolution of the perturbatively
calculable hard scattering $H_{f \bar f}$ with parton distributions
$\phi_{f_i/h}$, at factorization scale $\mu$, for parton $f_i$
carrying a momentum fraction $x_i$ of hadron $h$: 
\beqa
\frac{d\sigma_{h_a h_b\rightarrow Q{\bar Q}}}
{dQ^2 \, dy \, d\cos\theta}&=&\sum_{f \bar f} 
\int \frac{dx_a}{x_a} \frac{dx_b}{x_b} \, \phi_{f/h_a}(x_a,\mu^2) 
\nonumber \\ &&\hspace{-27mm}\times \, \phi_{{\bar f}/h_b}(x_b,\mu^2)
\; H_{f{\bar f}}\left(\frac{Q^2}{x_a x_b S},y,\theta,{Q\over \mu}\right),
\label{convolutionHQ}
\eeqa
where $S=(p_a+p_b)^2$, $Q^2=(p_1+p_2)^2$, and we sum over the two
main production partonic subprocesses, $q{\bar q} \rightarrow Q {\bar Q}$ 
and $gg \rightarrow Q {\bar Q}$.
We note that the short-distance hard scattering 
$H_{f {\bar f}}$ is a smooth function
only away from the edges of partonic phase space.
  
By replacing the incoming hadrons by partons in Eq.~(\ref{convolutionHQ}), 
we may write the infrared regularized partonic scattering 
cross section, after integrating over the total rapidity of the 
heavy quark pair, as
\beqa
\frac{d\sigma_{f{\bar f}\rightarrow Q{\bar Q}}}
{dQ^2 \; d\cos \theta}
&=&  \int_\tau^1 dz \, 
\int \frac {dx_a}{x_a} \, \frac{dx_b}{x_b} \, 
\phi_{f/f}(x_a,\mu^2,\epsilon) \, 
\nonumber \\ &&  \hspace{-5mm} \times \; 
\phi_{{\bar f}/{\bar f}}(x_b,\mu^2,\epsilon)
\quad \delta\left (z-{Q^2\over x_ax_bS}\right) \,  
\nonumber \\ && \hspace{-10mm} \times \,
{\hat \sigma}_{f{\bar f}\rightarrow Q{\bar Q}}\left(1-z, 
\frac{Q}{\mu},\theta,\alpha_s(\mu^2)\right),
\label{convpart}
\eeqa
where the argument $\epsilon$ represents the universal collinear singularities,
and we have introduced a simplified hard-scattering function, 
${\hat \sigma}_{f{\bar f}\rightarrow Q{\bar Q}}$. 
The threshold for the partonic subprocess is given 
in terms of the variable $z$,
\beq
z\equiv \frac{Q^2}{s} \, , 
\eeq 
with $s=x_a x_b S$ the invariant mass squared of the incoming partons.
At partonic threshold, $z_{\rm max}=1$, there is just enough 
partonic energy to produce the observed final state.
We also define a variable $\tau\equiv z_{\rm min}=Q^2/S$.
In general, ${\hat \sigma}$ includes ``plus''
distributions with respect to $1-z$, which arise from incomplete
cancellations near threshold between diagrams with
real gluon emission  and with virtual gluon corrections,
with singularities at $n$th order in $\alpha_s$ of the type
\beq
-\frac{\alpha_s^n}{n!}\, \left[\frac{\ln^{m}(1-z)}{1-z} \right]_{+}, \;  
m\le 2n-1\, .
\eeq

If we take Mellin transforms (moments with respect to the variable $\tau$)
of Eq.~(\ref{convpart}), the convolution
becomes a simple product of the moments of the parton distributions
and the hard scattering function ${\hat \sigma}$:
\beqa
&& \hspace{-10mm}\int_0^1 d\tau \, \tau^{N-1}
\frac{d\sigma_{f{\bar f}\rightarrow Q{\bar Q}}}
{dQ^2 \; d\cos \theta}={\tilde \phi}_{f/f}(N,\mu^2,\epsilon)
\nonumber \\ && \hspace{-10mm}\times  
{\tilde \phi}_{{\bar f}/{\bar f}}(N,\mu^2,\epsilon) \;
{\hat \sigma}_{f{\bar f}\rightarrow Q{\bar Q}}(N,
Q/\mu,\theta,\alpha_s(\mu^2))
\label{sigmom}
\eeqa
with the moments of the hard-scattering function and the parton
distributions defined, respectively,  by
\beqa 
\tilde{\sigma}(N)=\int_0^1dz\; z^{N-1}\hat{\sigma}(z) \, , 
\nonumber \\
\tilde{\phi}(N)=\int_0^1dx\; x^{N-1}\phi(x) \, .  
\label{momphisigma}
\eeqa
We then factorize the initial-state collinear divergences
into the parton distribution functions, expanded to the same
order in $\alpha_s$ as the partonic cross section, 
and thus obtain the perturbative expansion for the
infrared-safe hard scattering function, ${\hat \sigma}$.

We note that under moments divergent distributions in $1-z$ produce 
powers of $\ln N$:
\beqa
\int_0^1 dz\, z^{N-1}\left[{\ln^m(1-z)\over 1-z}\right]_+
&=&{(-1)^{m+1}\over m+1}\ln^{m+1}N 
\nonumber \\ && \hspace{-10mm}
+{\cal O}\left(\ln^{m-1}N\right)\, .
\eeqa

The hard scattering function ${\hat \sigma}$ is  sensitive
to soft-gluon dynamics through its $1-z$ dependence 
(or the $N$ dependence of its moments).   
We may now refactorize (moments of) the cross section 
into $N$-independent hard components $H_{IL}$, 
which describe the truly short-distance hard scattering,
center-of-mass distributions $\psi$, associated with gluons
collinear to the incoming partons, and a soft gluon function
$S_{LI}$ associated with non-collinear soft gluons. 
Note that the mass of the heavy
quarks eliminates final state collinear singularities in heavy quark 
production. 
This factorization is shown in Fig.~1 
where $f{\bar f}$ represents a pair of light quarks
or gluons that produce a heavy quark pair.
Fig. 2 defines the soft gluon function, $S_{LI}$,
which describes the coupling of soft gluons to the partons,
represented by eikonal lines, in the hard scattering.
$I$ and $L$ are color indices that describe the color structure
of the hard scattering.
The hard-scattering function takes contributions from the 
amplitude and its complex conjugate,
$H_{IL} ={h^*}_{L}\;h_{I}$.

\EPSFIGURE{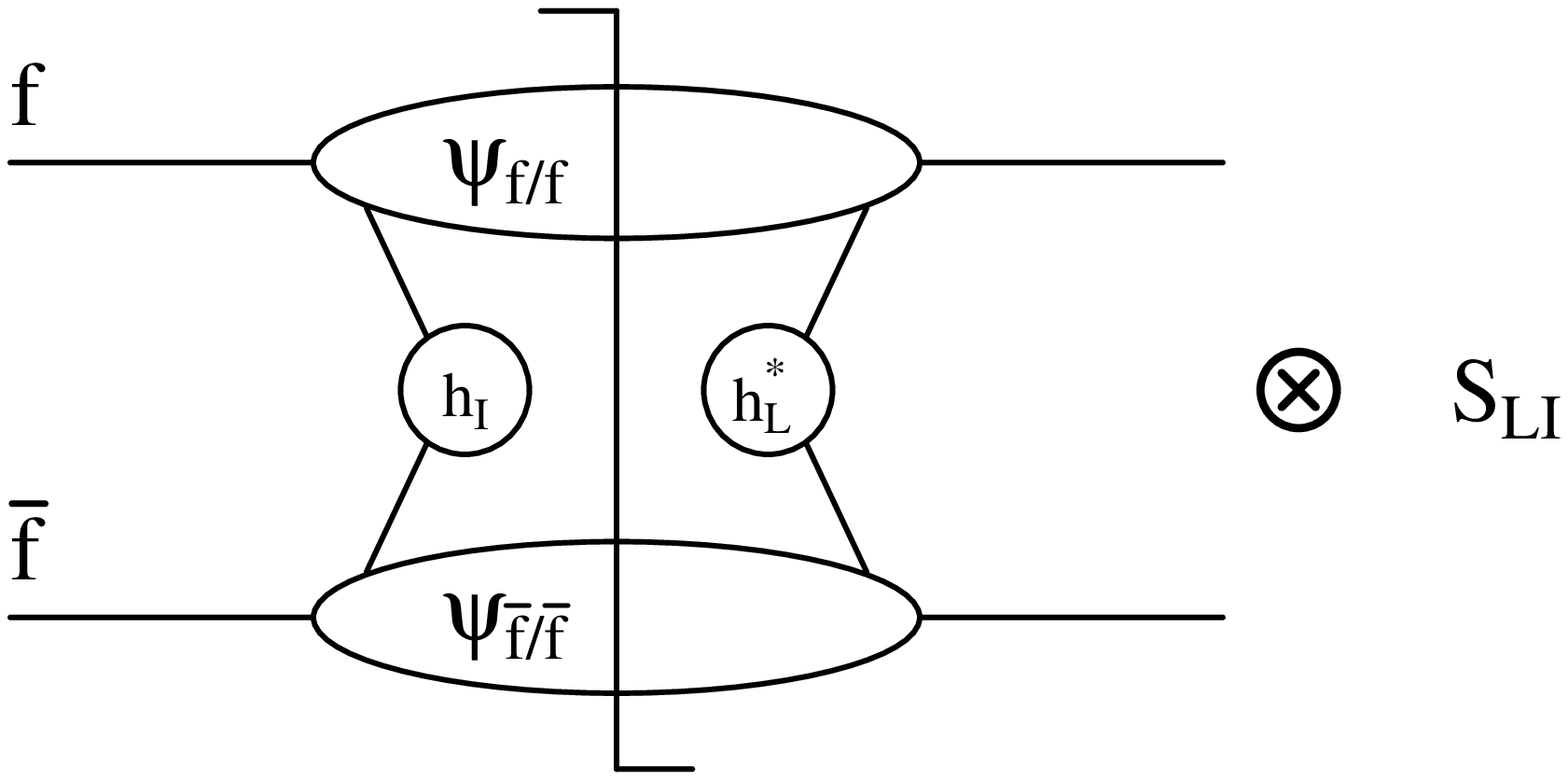,width=4.05in}
{Factorization for heavy quark production near 
partonic threshold.}

\EPSFIGURE{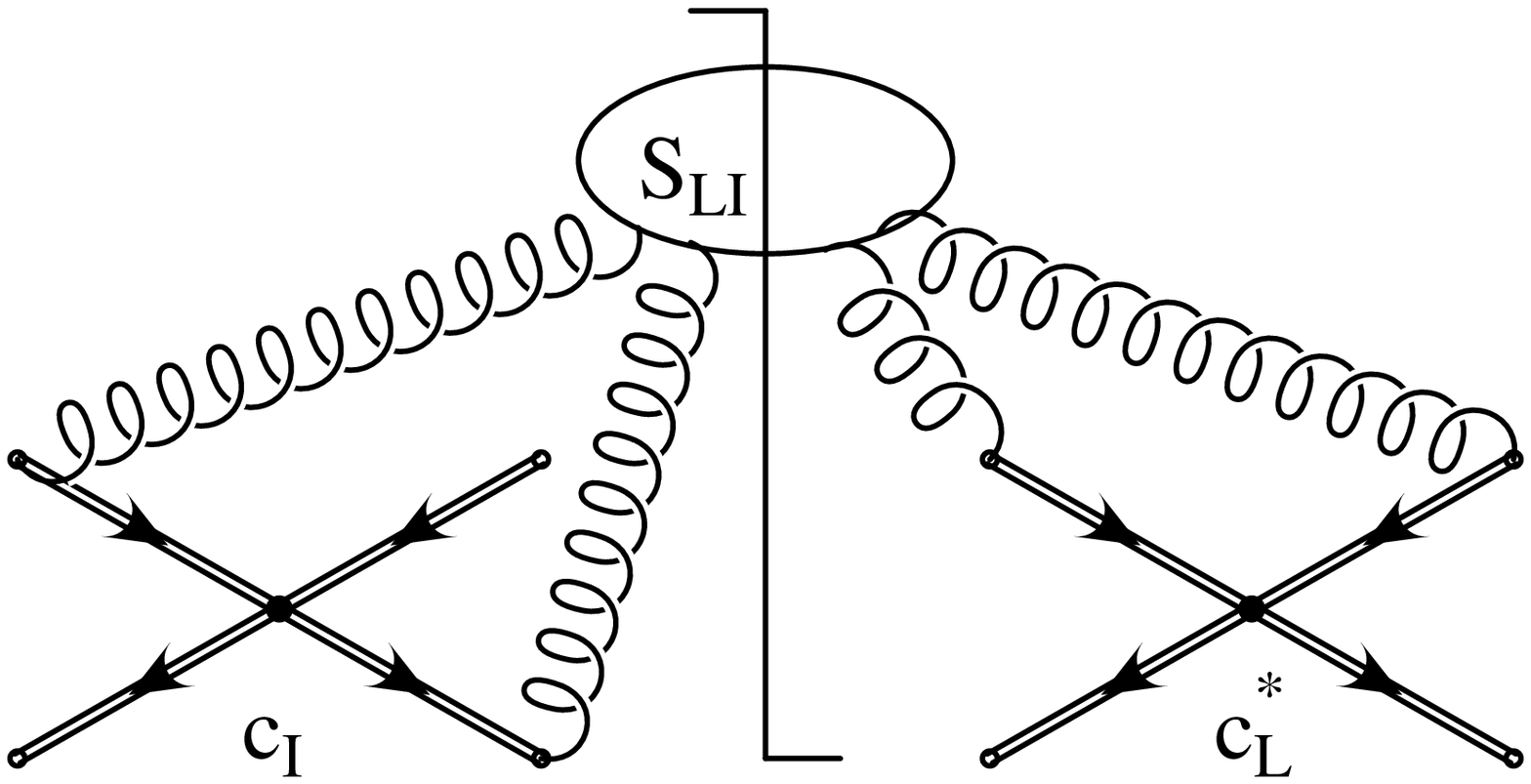,width=4.05in}
{The soft-gluon function $S_{LI}$ in which
the vertices $c_I,c_L^*$ link eikonal lines that represent
the incoming and outgoing partons.}

Then we write the refactorized cross section as~\cite{KS}
\beqa
&& \hspace{-4mm} \int_0^1 d\tau \tau^{N-1}
\frac{d\sigma_{f{\bar f}\rightarrow Q{\bar Q}}}
{dQ^2 \, d\cos \theta}
=\sum_{IL} H_{IL}\left({Q\over\mu},\theta, \alpha_s(\mu^2)\right) 
\nonumber \\ && \quad \quad \times \; 
{\tilde S}_{LI} \left({Q\over N \mu },\theta, \alpha_s(\mu^2) \right)
\nonumber \\ && \quad \quad \times \;
{\tilde\psi}_{f/f}\left ( N,{Q\over \mu },\epsilon \right) \;
{\tilde\psi}_{{\bar f}/{\bar f}}\left (N,{Q\over \mu },\epsilon \right ) .
\label{sigref}
\eeqa

The center-of-mass distribution functions $\psi$ absorb 
the universal collinear singularities
of the initial-state partons in the refactorized cross section. 
They differ from the standard light-cone parton 
distributions $\phi$ by being defined at fixed energy rather than
light-like momentum fraction.  
The moments of $\psi$ can then be written as products of moments
of $\phi$ and an infrared safe function.

\subsection{Resummation}

Now, comparing Eqs.\ (\ref{sigmom}) and (\ref{sigref}), we
see that the moments of the partonic
heavy-quark production cross section are given by 
\beqa
&&{\tilde \sigma}_{f{\bar f}\rightarrow Q{\bar Q}}(N)=
\left[{{\tilde\psi}_{f/f}(N,Q/\mu,\epsilon)
\over{\tilde \phi}_{f/f}(N,\mu^2,\epsilon)}\right]^2\,
\nonumber \\ && \hspace{-2mm} \times
\sum_{IL} H_{IL}\left(\frac{Q}{\mu},\theta,\alpha_s(\mu^2)\right)
{\tilde S}_{LI}\left(\frac{Q}{N\mu},\theta,\alpha_s(\mu^2)\right) 
\nonumber \\ &&
\label{psiphiHS}
\eeqa
where $f{\bar f}$ denotes $q{\bar q}$ or $gg$,
and we have used the relations
${\psi}_{q/q}={\psi}_{{\bar q}/{\bar q}}$
and ${\phi}_{q/q}={\phi}_{{\bar q}/{\bar q}}$.
We will now discuss the exponentiation of logarithms of $N$
for each factor in the above equation.

The first factor, $(\tilde \psi_{f/f}/\tilde \phi_{f/f})^2$, 
in Eq.~(\ref{psiphiHS})
is universal between electroweak and QCD
hard processes, and was computed first for
the Drell-Yan cross section \cite{St87}.  
The resummed expression for this factor is
\beqa
&&\frac{{\tilde{\psi}}_{f/f}(N,Q/\mu,\epsilon)}
{{\tilde{\phi}}_{f/f}(N,\mu^2,\epsilon)}
= R_{(f)}\, \exp \left[E^{(f)}(N,Q^2)\right]
\nonumber\\ &&  \quad \times
\exp \left\{-2\int_{\mu}^Q \frac{d\mu'}{\mu'}\; 
\left [\gamma_f(\alpha_s(\mu'{}^2))\right. \right.
\nonumber\\ && \hspace{25mm} \left. \left.
-\gamma_{ff}(N,\alpha_s(\mu'{}^2)) \right]\right\}\, ,
\label{psiphimu}
\eeqa
where $\gamma_f$ is the anomalous dimension of the field of flavor $f$,
which is independent of $N$, and 
$\gamma_{ff}$ is the anomalous dimension of the color-diagonal 
splitting function for flavor $f$.
$R_{(f)}(\alpha_s)$ is an $N$-independent function of the coupling, 
which can be  normalized to unity at zeroth order.
The exponent $E^{(f)}$ is given by
\beqa
&& \hspace{-8mm} E^{(f)}\left(N,Q^2\right)=
-\int^1_0 dz \frac{z^{N-1}-1}{1-z}
\nonumber \\ && \hspace{-3mm} \times  
\left \{\int^{(1-z)^{m_S}}_{(1-z)^2} \frac{d\lambda}{\lambda} 
A^{(f)}\left[\alpha_s(\lambda Q^2)\right] \right.
\nonumber\\ && \quad
{}+B^{(f)}\left[\alpha_s((1-z)^{m_s} Q^2)\right]
\nonumber\\ && \quad \left.
{}+\frac{1}{2}\nu^{(f)}\left[\alpha_s((1-z)^2 Q^2)\right] \right\} \, ,
\label{Eexp}
\eeqa
where the parameter $m_S$ and the resummed coefficients $B^{(f)}$ 
depend on the factorization scheme. 
In the  DIS and $\overline{\rm MS}$ factorization schemes
we have $m_S=1$ and $m_S=0$, respectively.
The function $A^{(f)}$ is given by 
\beq
A^{(f)}(\alpha_s) = C_f\left ( {\alpha_s\over \pi} 
+\frac{1}{2} K \left({\alpha_s\over \pi}\right)^2\right )\, ,
\label{Aexp}
\eeq
with $C_f=C_F=(N_c^2-1)/(2N_c)$ for an incoming quark, 
and $C_f=C_A=N_c$ for an incoming gluon, with $N_c$ the number of colors,
and $K=C_A(67/18-\pi^2/6)-5n_f/9$,
where $n_f$ is the number of quark flavors.  
$B^{(f)}$ is given for quarks in the DIS scheme by
$B^{(q)}(\alpha_s)=-3C_F\alpha_s/(4\pi)$, 
while it vanishes in the $\overline {\rm MS}$ scheme for quarks and gluons.  
Finally, the lowest-order approximation to  
the scheme-independent $\nu^{(f)}$ is 
$\nu^{(f)}=2C_f \alpha_s/\pi$.

Next, we discuss resummation for the soft function.
The soft matrix $S_{LI}$ depends on $N$ through the ratio $Q/(N\mu)$;
its $N$-dependence can then be resummed by renormalization group
analysis. The product $H_{IL}S_{LI}$
of the soft function and the hard factors needs
no overall renormalization, because the UV divergences of $S_{LI}$
are balanced by construction by those of $H_{IL}$.
Thus, we have~\cite{KS}
\beqa
H^{(0)}_{IL}&=& \prod_{i=f,\bar f} Z_i^{-1}\; \left(Z_S^{-1}\right)_{IC}
H_{CD} \left[\left(Z_S^\dagger \right)^{-1}\right]_{DL} \, ,
\nonumber \\ 
S^{(0)}_{LI}&=&(Z_S^\dagger)_{LB}S_{BA}Z_{S,AI},
\label{HSren}
\eeqa
where $H^{(0)}$ and $S^{(0)}$ denote the unrenormalized quantities,
$Z_i$ is the renormalization constant of the $i$th
incoming partonic field external to $h_I$, and $Z_{S,LI}$ is
a matrix of renormalization constants which describes the
renormalization of the soft function, including
mixing of color structures, and which is defined to include the
wave function renormalization for the heavy quarks.

From Eq.\ (\ref{HSren}), the soft function  $S_{LI}$ satisfies the
renormalization group equation 
\beqa
\left(\mu {\partial \over \partial \mu}+\beta(g){\partial \over \partial g}
\right)\,S_{LI}&=&-(\Gamma^\dagger_S)_{LB}S_{BI}
\nonumber \\ && \hspace{-8mm}
-S_{LA}(\Gamma_S)_{AI}\, ,
\label{RGE}
\eeqa
where $\Gamma_S$ is an anomalous dimension matrix 
that is calculated by explicit renormalization of the soft function.
In a minimal subtraction renormalization scheme and with
$\epsilon=4-n$, where $n$ is the number of space-time dimensions,
the matrix of anomalous dimensions at one loop is given by
\begin{equation}
\Gamma_S (g)=-\frac{g}{2} \frac {\partial}{\partial g}{\rm Res}_{\epsilon
\rightarrow 0} Z_S (g, \epsilon) \, .
\end{equation}

Using Eqs.~(\ref{psiphiHS}), (\ref{psiphimu}), and the solution of the
renormalization group equation for the soft function, we can write 
the resummed heavy quark cross section in moment space as 
\beqa
&& \hspace{-3mm}\tilde{{\sigma}}_{f{\bar f}\rightarrow Q{\bar Q}}(N)
=R_{(f)}^2 \; \exp \left \{2 \left[ E^{(f_i)}(N,Q^2) \right. \right. 
\nonumber\\&& \hspace{-5mm} \left. \left.
{}-2\int_\mu^Q{d\mu'\over\mu'} 
\left[\gamma_{f_i}(\alpha_s(\mu'{}^2))-\gamma_{f_if_i}(N,\alpha_s(\mu'{}^2)) 
\right] \right] \right\}
\nonumber\\ && \hspace{-3mm} \quad \times \; {\rm Tr} \left \{
H\left({Q\over\mu},\theta,\alpha_s(\mu^2)\right) \right. 
\nonumber\\ && \hspace{-3mm} \quad \times \;
\bar{P} \exp \left[\int_\mu^{Q/N} {d\mu' \over \mu'} \;
\Gamma_S^\dagger\left(\alpha_s(\mu'^2)\right)\right] 
\nonumber\\ &&  \hspace{-3mm} \quad \times \; 
{\tilde S} \left(1,\theta,\alpha_s(Q^2/N^2) \right) \; 
\nonumber\\ && \hspace{-3mm} \quad \left. \times \;
P \exp \left[\int_\mu^{Q/N} {d\mu' \over \mu'}\; \Gamma_S
\left(\alpha_s(\mu'^2)\right)\right] \right\}\, ,
\label{resHQ}
\eeqa
where the symbols $P$ and $\bar{P}$ refer to path ordering
and the trace is taken in color space.

We may simplify this result by choosing a color basis in which 
the anomalous dimension matrix $\Gamma_S$ is diagonal, 
with eigenvalues $\lambda_I$ for each basis color
tensor labelled by $I$.
Then, we have
\begin{eqnarray}
&& \hspace{-5mm}
{\tilde S}_{LI}\left(\frac{Q}{N\mu}, \theta, \alpha_s(\mu^2)\right)=
{\tilde S}_{LI}\left(1,\theta, \alpha_s\left(\frac{Q^2}{N^2}\right)\right)
\nonumber \\ && \hspace{-5mm} \times \, 
\exp\left[-\int^{\mu}_{Q/N}\frac{d \bar{\mu}}{\bar{\mu}}
[\lambda_I(\alpha_s(\bar{\mu}^2))
+\lambda^*_L(\alpha_s(\bar{\mu}^2))]\right]\, .
\nonumber \\
\label{rgedgsol}
\end{eqnarray}
Thus, in a diagonal basis, and with $\mu=Q$ and $R_{(f)}$ normalized to unity,
we can rewrite the resummed cross section in a simplified form as 
\beqa
&& \hspace{-3mm}\tilde{{\sigma}}_{f{\bar f}\rightarrow Q{\bar Q}}(N) = 
H_{IL}\left(1,\theta,\alpha_s(\mu^2)\right) 
\nonumber \\ && \hspace{-3mm}
\times {\tilde S}_{LI} \left(1,\theta,\alpha_s(Q^2/N^2) \right) \, 
\exp \left[E_{LI}^{(f {\bar f})}(N,\theta,Q^2)\right] \, ,
\nonumber \\ &&
\label{crossdiag}
\eeqa
where the exponent is
\begin{eqnarray}
&& \hspace{-5mm}
E_{LI}^{(f {\bar f})}(N,\theta,Q^2)=-\int_0^1 dz \frac{z^{N-1}-1}{1-z}
\nonumber \\ &&  \times \, 
\left\{\int^{(1-z)^{m_S}}_{(1-z)^2} \frac{d\lambda}{\lambda}
g_1^{(f{\bar f})}[\alpha_s(\lambda Q^2)] \right.
\nonumber \\ &&  \quad \quad
{}+g_2^{(f{\bar f})}[\alpha_s((1-z)^{m_S} Q^2)]
\nonumber \\ && \quad \quad \left.
{}+g_3^{(IL,f{\bar f})}[\alpha_s((1-z)^2 Q^2),\theta] \right\}\, ,
\nonumber \\
\label{ELI}
\end{eqnarray}
with the functions $g_1$, $g_2$, and $g_3$ defined by
\beqa
&&g_1^{(f\bar f)}=A^{(f)}+A^{(\bar f)} \, , 
\quad g_2^{(f \bar f)}=B^{(f)}+B^{(\bar f)} \, ,
\nonumber \\ &&
g_3^{(IL,f {\bar f})}=-\lambda_I-\lambda_L^*+\frac{1}{2}\nu^{(f)}
+\frac{1}{2}\nu^{(\bar f)} \, .
\label{g1g2g3}
\eeqa

In the next section we present the soft anomalous dimension matrices
for heavy quark production through light quark annihilation and gluon fusion;
also we give NLO expansions of the resummed cross section in both
partonic production channels.

\section{Soft anomalous dimension matrices for heavy quark production}

\subsection{Soft anomalous dimension for $q {\bar q} \rightarrow Q {\bar Q}$}

First, we present the soft anomalous dimension matrix for heavy quark
production through light quark annihilation,
\begin{equation}
q(p_a,r_a)+{\bar q}(p_b,r_b) \rightarrow {\bar Q}(p_1,r_1) + Q(p_2,r_2)\, ,
\end{equation}
where the $p_i$'s and $r_i$'s denote momenta and colors of the partons
in the process.

We introduce the Mandelstam invariants
\beqa
s&=&(p_a+p_b)^2 \, , 
\nonumber \\
t_1&=&(p_a-p_1)^2-m^2 \, , 
\nonumber \\ 
u_1&=&(p_b-p_1)^2-m^2 \, ,
\label{Mandelstam}
\eeqa
with $m$ the heavy quark mass, which satisfy
$s+t_1+u_1=0$ at partonic threshold.

\EPSFIGURE{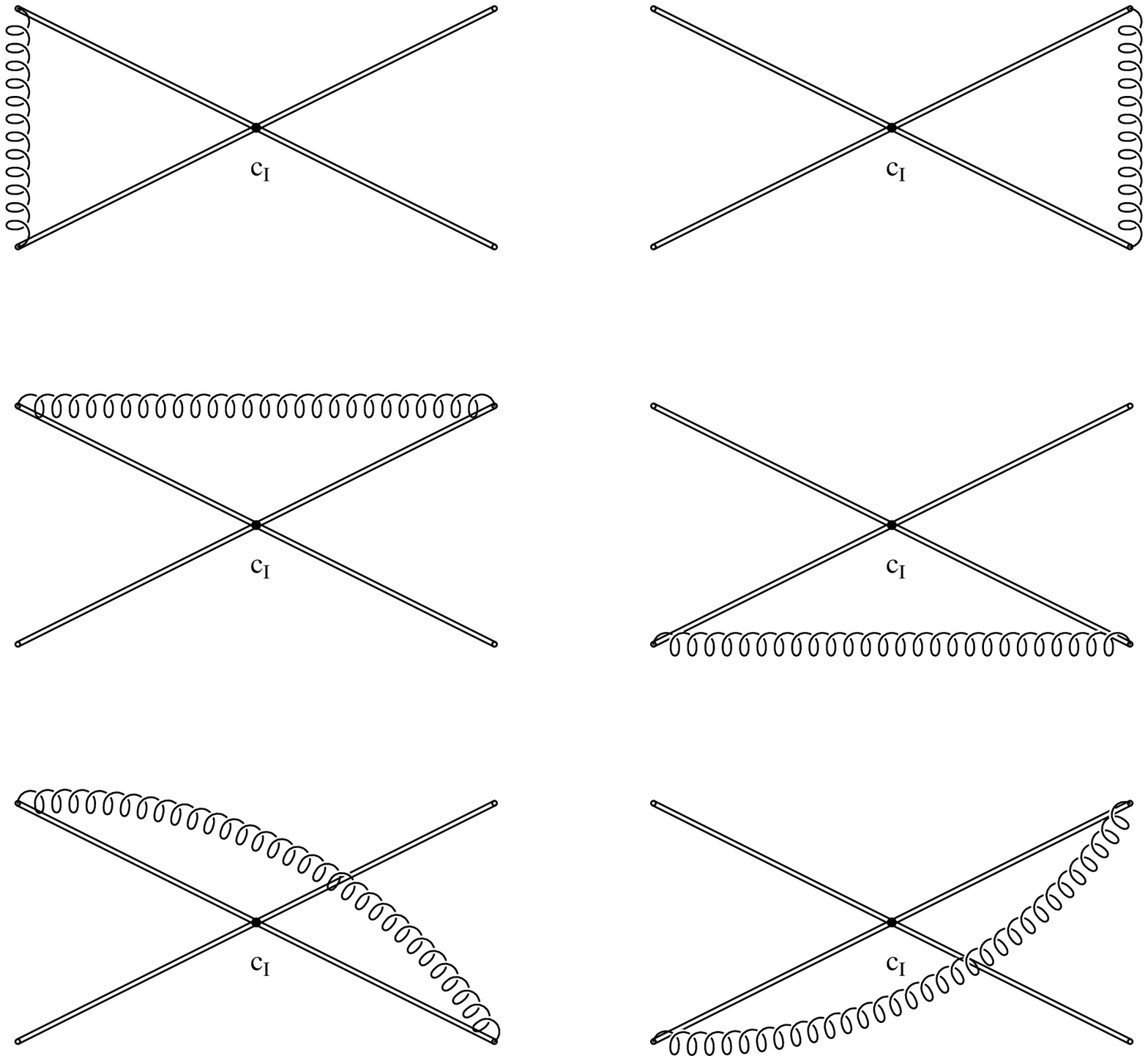,width=4.05in}
{Eikonal vertex corrections to $S_{LI}$ 
for partonic subprocesses in heavy quark or dijet production.}
\EPSFIGURE{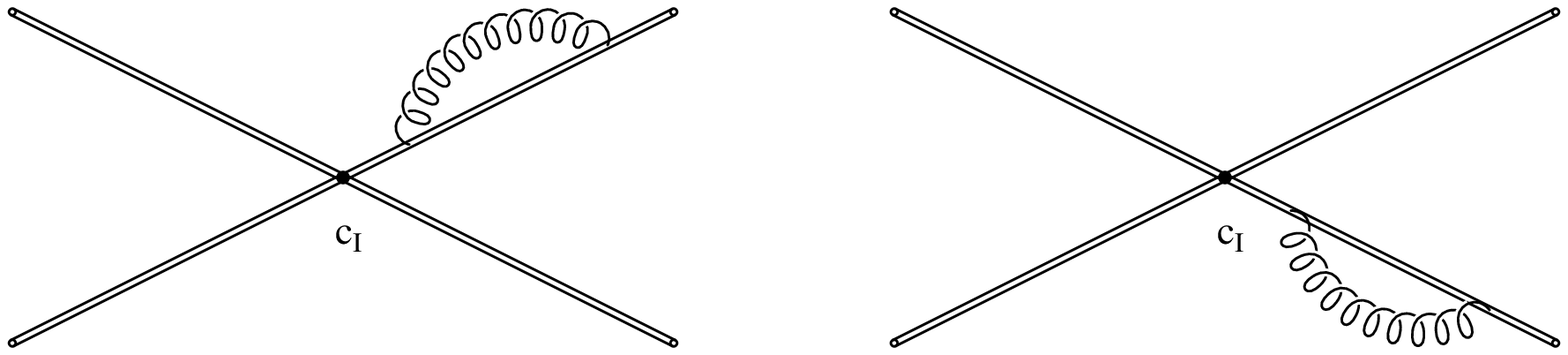,width=4.05in}
{Eikonal self-energy graphs for heavy quark production.}
 
For the determination of $\Gamma_S$ we evaluate the one-loop graphs 
in Figs. 3 and 4.
For the $q {\bar q} \rightarrow Q {\bar Q}$ partonic subprocess
$\Gamma_S$ is a $2 \times 2$ matrix, since the process can be described
by a basis of two color tensors.
Here we calculate the soft matrix 
in the $s$-channel singlet-octet basis:
\beqa
c_1 &=& c_{\rm singlet}=\delta_{r_ar_b}\delta_{r_1r_2}\, , \quad \quad
\nonumber \\ 
c_2 &=& c_{\rm octet}=(T^c_F)_{r_b r_a}(T^c_F)_{r_2 r_1}\, ,
\label{18basis}
\eeqa
where the $T_F^c$ are the generators of $SU(3)$ in the fundamental 
representation.
The result of our calculation in a general axial gauge is~\cite{NK,Thesis,KS} 
\begin{equation}
\Gamma_S=\left[\begin{array}{cc}
\Gamma_{11} & \Gamma_{12} \\
\Gamma_{21} & \Gamma_{22}
\end{array}
\right] \, ,
\label{matrixqqQQ}
\end{equation}
with 
\begin{eqnarray}
\Gamma_{11}&=&-\frac{\alpha_s}{\pi}C_F \, [L_{\beta}
+\ln(2\sqrt{\nu_a\nu_b})+\pi i] \, ,
\nonumber \\
\Gamma_{21}&=&\frac{2\alpha_s}{\pi}
\ln\left(\frac{u_1}{t_1}\right) \, ,
\nonumber \\
\Gamma_{12}&=&\frac{\alpha_s}{\pi}
\frac{C_F}{C_A} \ln\left(\frac{u_1}{t_1}\right) \, ,
\nonumber \\
\Gamma_{22}&=&\frac{\alpha_s}{\pi}
\nonumber \\ && \hspace{-10mm}  \times 
\left\{C_F\left[4\ln\left(\frac{u_1}{t_1}\right)
-\ln(2\sqrt{\nu_a\nu_b})
-L_{\beta}-\pi i\right]\right.
\nonumber \\ && \hspace{-12mm}
\left.{}+\frac{C_A}{2}\left[-3\ln\left(\frac{u_1}{t_1}\right)
-\ln\left(\frac{m^2s}{t_1u_1}\right)+L_{\beta}+\pi i \right]\right\}\, ,
\nonumber \\
\label{GammaqqQQ}
\end{eqnarray}
where $L_\beta$ is the  velocity-dependent
eikonal function
\begin{equation}
L_{\beta}=\frac{1-2m^2/s}{\beta}\left(\ln\frac{1-\beta}{1+\beta}
+\pi i \right)\, ,
\end{equation}
with $\beta=\sqrt{1-4m^2/s}$. 
The gauge dependence of the incoming eikonal lines is given in terms of   
$\nu_i \equiv (v_i \cdot n)^2/|n|^2$, with $n^{\mu}$ the gauge vector,
which cancels against corresponding terms in the parton distributions.
Here the dimensionless velocity vectors $v_i^{\mu}$ are defined
by $p_i^{\mu}=Q v_i^{\mu}/\sqrt{2}$ and obey $v_i^2=0$ for the
light incoming quarks and $v_i^2=2m^2/Q^2$ for the outgoing
heavy quarks. 
The gauge dependence of the outgoing heavy quarks is cancelled
by the inclusion of the self-energy diagrams in Fig. 4.

$\Gamma_S$ is diagonalized in this singlet-octet color basis
at absolute threshold, $\beta=0$. It is also diagonalized
for arbitrary $\beta$ when the parton-parton c.m. scattering angle is
$\theta=90^\circ$ (where $u_1=t_1$), with
eigenvalues that may be read off from Eq.\ (\ref{GammaqqQQ}).

We can expand the resummed heavy quark production
cross section to any fixed order
in perturbation theory without having to diagonalize the soft anomalous
dimension matrix. 

The NLO expansion for $q {\bar q} \rightarrow Q {\bar Q}$
in the $\overline{\rm MS}$ scheme is
\begin{eqnarray}
&& \hspace{-3mm}
{\hat \sigma}^{\overline{\rm MS} \, (1)}_{q{\bar q}\rightarrow Q{\bar Q}}
(1-z,m^2,s,t_1,u_1)=\sigma^B_{q{\bar q}\rightarrow Q{\bar Q}}
\frac{\alpha_s}{\pi}
\nonumber \\ && \hspace{-3mm} \times
\left\{4C_F\left[\frac{\ln(1-z)}{1-z}\right]_{+}
+\left[\frac{1}{1-z}\right]_{+} \right.
\nonumber \\ && \hspace{-3mm} \times 
\left[C_F\left(8\ln\left(\frac{u_1}{t_1}\right)
-2-2 \, {\rm Re} \, L_{\beta}+2\ln\left(\frac{s}{\mu^2}\right)\right)\right.
\nonumber \\ && \hspace{-5mm}
\left.\left.
{}+C_A\left(-3\ln\left(\frac{u_1}{t_1}\right)+{\rm Re} \, L_{\beta}
-\ln\left(\frac{m^2s}{t_1 u_1}\right)\right)\right]\right\}\, ,
\nonumber \\
\end{eqnarray}
where $\sigma^B_{q{\bar q}\rightarrow Q{\bar Q}}$ is the Born cross section.
Similar results have been obtained in the DIS scheme.
These results are consistent with the approximate one-loop results
in Ref.~\cite{mengetal}.
Analytical and numerical results have also been obtained for the expansions 
at NNLO \cite{NK,KLMV} in both pair inclusive and single-particle
inclusive kinematics.

\subsection{Soft anomalous dimension for $gg \rightarrow Q {\bar Q}$}

Next, we present the soft anomalous dimension matrix for heavy quark
production through gluon fusion,
\begin{equation}
g(p_a,r_a)+g(p_b,r_b) \rightarrow {\bar Q}(p_1,r_1) + Q(p_2,r_2)\, ,
\end{equation}
with momenta and colors labelled by the $p_i$'s and $r_i$'s, respectively. 
Here $\Gamma_S$ is a $3 \times 3$ matrix since the process can be described
by a basis of three color tensors.
The relevant graphs for the calculation of $\Gamma_S$
are the same as in Figs.~3 and 4, where now the incoming
eikonal lines represent gluons.

We make the following choice for the color basis:
\beqa 
c_1&=&\delta_{r_ar_b}\,\delta_{r_2r_1} \, , 
\nonumber \\ 
c_2&=&d^{r_ar_bc}\,(T^c_F)_{r_2r_1} \, ,
\nonumber \\ 
c_3&=&i f^{r_ar_bc}\,(T^c_F)_{r_2r_1} \, ,
\eeqa
where $d^{abc}$ and $f^{abc}$ are the totally
symmetric and antisymmetric $SU(3)$ invariant tensors, respectively.
Then the anomalous dimension matrix in a 
general axial gauge is given by~\cite{NK,Thesis,KS}
\begin{equation}
\Gamma_S=\left[\begin{array}{ccc}
\Gamma_{11} & 0 & \frac{\Gamma_{31}}{2} \vspace{2mm} \\
0 & \Gamma_{22} & \frac{N_c}{4} \Gamma_{31} \vspace{2mm} \\
\Gamma_{31} & \frac{N_c^2-4}{4N_c}\Gamma_{31} & \Gamma_{22}
\end{array}
\right] \, ,
\label{GammaggQQ33}
\end{equation}
where
\begin{eqnarray}
&& \Gamma_{11}=\frac{\alpha_s}{\pi}\left [-C_F(L_{\beta}+1) \right.
\nonumber \\ && \hspace{15mm} \left.
{}+C_A\left(-\frac{1}{2}\ln\left({4\nu_a\nu_b}\right) 
+1-\pi i\right)\right ],
\nonumber \\ &&
\Gamma_{31}=\frac{\alpha_s}{\pi}\ln\left(\frac{u_1^2}{t_1^2}\right) \, ,
\nonumber \\ &&
\Gamma_{22}=\frac{\alpha_s}{\pi}\left\{-C_F(L_{\beta}+1)
+\frac{C_A}{2} \right.
\nonumber \\ && \left. \times 
\left[-\ln\left(4\nu_a\nu_b\right) 
+2+\ln\left(\frac{t_1 u_1}{m^2 s}\right)+L_{\beta}-\pi i
\right]\right\}.
\nonumber \\
\label{GammaggQQ}
\end{eqnarray}
We note that $\Gamma_S$ is diagonalized in this basis
at absolute threshold, $\beta=0$, and also for arbitrary $\beta$ 
when the parton-parton c.m. scattering angle is
$\theta=90^\circ$ , with
eigenvalues that may be read off from Eq.\ (\ref{GammaggQQ}).

Again, we may expand the resummed cross section for 
$gg \rightarrow Q {\bar Q}$ at NLO and NNLO or higher.
The NLO expansion in the ${\overline{\rm MS}}$ scheme is 
\beqa
&& \hspace{-3mm} 
{\hat \sigma}^{\overline {\rm MS} \, (1)}_{gg \rightarrow Q{\bar Q}}
(1-z,m^2,s,t_1,u_1)=
\sigma^B_{gg\rightarrow Q{\bar Q}} \frac{\alpha_s}{\pi} 
\nonumber \\ && \hspace{-4mm}\times 
\left\{4C_A\left[\frac{\ln(1-z)}{1-z}\right]_{+}
-2C_A \ln\left(\frac{\mu^2}{s}\right) 
\left[\frac{1}{1-z}\right]_{+}\right\}
\nonumber \\ && \hspace{-4mm}
{}+\alpha_s^3 K_{gg} B_{QED} \left[\frac{1}{1-z}\right]_{+}
\left\{N_c(N_c^2-1)\frac{(t_1^2+u_1^2)}{s^2} \right.
\nonumber \\ && \hspace{-4mm} \times 
\left[\left(-C_F+\frac{C_A}{2}\right)
{\rm Re} \, L_{\beta}
+\frac{C_A}{2}\ln\left(\frac{t_1u_1}{m^2s}\right)
-C_F\right]
\nonumber \\ && \hspace{-4mm}
{}+\frac{N_c^2-1}{N_c}(C_F-C_A) {\rm Re} \, L_{\beta}
-(N_c^2-1)\ln\left(\frac{t_1u_1}{m^2s}\right)
\nonumber \\ && \hspace{-4mm} \left.
{}+C_F \frac{N_c^2-1}{N_c}
+\frac{N_c^2}{2}(N_c^2-1)
\ln\left(\frac{u_1}{t_1}\right)\frac{(t_1^2-u_1^2)}{s^2} \right\} 
\nonumber\\&&
\label{gg1loop}
\eeqa
where $\sigma^B_{gg\rightarrow Q{\bar Q}}$ is the Born cross section,
\beq
B_{\rm QED}=\frac{t_1}{u_1}+\frac{u_1}{t_1}+\frac{4m^2s}{t_1u_1}
\left(1-\frac{m^2s}{t_1u_1}\right), 
\eeq
and $K_{gg}=(N^2-1)^{-2}$ is a color average factor.

\section{Diagonalization and numerical results for
$q {\bar q} \rightarrow t{\bar t}$}

In this section we give some numerical results for top quark
production at the Tevatron.

As we noted in the previous section, the soft anomalous
dimension matrices, $\Gamma_S$,  are in general not diagonal. 
They are only diagonal at specific kinematical regions, i.e.
at absolute threshold, $\beta=0$, 
and at a scattering angle $\theta=90^{\circ}$. In these cases the
exponentiated cross section has a simpler form, Eq.~(\ref{crossdiag}),
and numerical studies are easier to perform. Thus,
the resummed cross section at $\theta=90^{\circ}$ for top quark 
production at the Fermilab Tevatron and bottom quark production
at HERA-B were presented in Ref.~\cite{NKJSRV}.

In general, however, we must diagonalize $\Gamma_S$, i.e. we must 
find new color bases where
$\Gamma_S$ is diagonal so that the resummed cross section 
can take the  simpler form of Eq. (\ref{crossdiag}).
The diagonal basis is given by $C'=CR$ where the columns of the matrix
$R$ are the eigenvectors of $\Gamma_S$.
A detailed discussion of the diagonalization procedure
has been given in Refs.~\cite{NK,NKRV}.

The resummed partonic cross section for the process
$q {\bar q} \rightarrow Q {\bar Q}$
can be written as ~\cite{NKRV}
\beqa && \hspace{-4mm}
\sigma^{\rm res}_{q \bar q}(s,m^2)=\sum_{i,j=1}^2\int_{-1}^1 d\cos \theta \,
\nonumber \\ && \hspace{-4mm} \times 
\left[-\int^{s-2ms^{1/2}}_{s_{\rm cut}}
ds_4 \, f_{q \bar q, ij}(s_4, \theta)
\frac{d{\overline \sigma}_{q \bar q, ij}^{(0)}(s,s_4,\theta)}{ds_4}\right] 
\nonumber \\ 
\label{respart}
\eeqa
where $s_4=s+t_1+u_1$.
The $d{\overline \sigma}_{q \bar q, ij}^{(0)}(s,s_4,\theta)/ds_4$
are color components of the differential of the Born cross section.
The function $f_{q \bar q, ij}$ is given at NLL by the exponential
\beq
f_{q \bar q, ij}=\exp[E_{q \bar q, ij}]
=\exp[E_{q \bar q}+E_{q \bar q}(\lambda_i, \lambda_j)] \, ,
\eeq
where in the DIS scheme
\beqa &&
E_{q\overline q}^{\rm DIS} = \int_{\omega_0}^1\frac{d\omega'}{\omega'}
\Big\{\int_{\omega'^2 Q^2/\Lambda^2}^{\omega' Q^2/\Lambda^2} \frac{d\xi}{\xi}\,
 \Big[ \frac{2 C_F}{\pi} \Big( \alpha_s(\xi)
\nonumber \\ && \quad
{}+ \frac{1}{2\pi} \alpha^2_s(\xi) K\Big) \Big]
 - \frac{3}{2} \frac{C_F}{\pi} \alpha_s
\Big( \frac{\omega' Q^2}{\Lambda^2}\Big)
\, \Big\} \, ,
\label{Eofm}
\eeqa
with $\omega_0=s_4/(2m^2)$ and $\Lambda$ the QCD scale parameter.
The color-dependent contribution in the exponent is
\beqa
E_{q\overline q}(\lambda_i, \lambda_j)&=&
-\int_{\omega_0}^1\frac{d\omega'}{\omega'}
\left\{\lambda_i' \left[\alpha_s \left(\frac{\omega'^2 Q^2}
{\Lambda^2}\right), \theta \right] \right.
\nonumber \\ && \hspace{-2mm} \left.
+{\lambda_j'}^* \left[\alpha_s \left(\frac{\omega'^2 Q^2}{\Lambda^2}\right),
\theta \right] \right\} \, ,
\eeqa
in both mass factorization schemes, where $i,j= 1,2$.
Here $\lambda'=\lambda-\nu^{(f)}/2$ (see Eq.~(\ref{g1g2g3})) where
the $\lambda$'s are the eigenvalues of the soft anomalous dimension matrix.
The cutoff $s_{\rm cut}$ in Eq.~(\ref{respart}) 
regulates the divergence of $\alpha_s$ at low
$s_4$: $s_{\rm cut}>s_{4,{\rm min}}=2m^2\Lambda/Q$.
We choose a value of the cutoff consistent with the sum of the 
first few terms in the perturbative expansion \cite{NKJSRV,LSN,HERAB}, 
in the range $30 s_{4, {\rm min}} < s_{\rm cut} < 40 s_{4, {\rm min}}$,
which corresponds to a soft gluon energy cutoff of the order
of the decay width of the top, thus giving a natural boundary
of the nonperturbative region (see also the discussion in \cite{BC}).
The central value in our cutoff range, $s_{\rm cut} = 35 s_{4, {\rm min}}$,
corresponds to $s_4/(2m^2)=0.04$ for $m=175$ GeV/$c^2$
and $\Lambda=0.2$ GeV.

After the diagonalization of the soft anomalous dimension matrix,
the resummed partonic cross section is given at NLL accuracy by \cite{NKRV}
\beqa 
&&  
\sigma^{\rm res}_{q \overline q}(s,m^2) 
= -\int_{-1}^1 d\cos \theta \, 
\int^{s-2ms^{1/2}}_{s_{\rm cut}} ds_4 \;
\nonumber \\ && \times 
\frac{1}{|\lambda_1-\lambda_2|^2} 
\frac{d{\overline \sigma}_{q \overline q}^{(0)}(s,s_4,\theta)}{ds_4}
\nonumber \\ &&  \times
\left[\left(\frac{4N_c^2}{N_c^2-1}\Gamma_{12}^2 
+|\lambda_1-\Gamma_{11}|^2 \right) e^{E_{q \overline q, 11}} \right. 
\nonumber \\ &&
{}+\left(\frac{4N_c^2}{N_c^2-1}\Gamma_{12}^2 
+|\lambda_2-\Gamma_{11}|^2 \right) e^{E_{q \overline q, 22}}
\nonumber \\ && 
{}-\frac{8N_c^2}{N_c^2-1}\Gamma_{12}^2 
{\rm Re}\left(e^{E_{q \overline q, 12}}\right)
\nonumber \\ &&  \left.
{}-2{\rm Re} \left((\lambda_1-\Gamma_{11})(\lambda_2-\Gamma_{11})^*
e^{E_{q \overline q, 12}}\right)\right].
\label{partres}
\eeqa

To calculate the NLL resummed hadronic cross section we convolute
the parton distributions $\phi_{i/h}$, for parton $i$ in hadron $h$,
with the partonic cross section: 
\begin{eqnarray} &&
\sigma^{\rm res}_{q \overline q, {\rm had}}(S,m^2)=\sum_{q=u}^b
\int_{\tau_0}^1 d\tau
\int_\tau^1 \frac{dx}{x}
\phi_{q/h_1}(x,\mu^2) 
\nonumber \\ && \quad \quad \times \, 
\phi_{{\bar q}/h_2}(\frac{\tau}{x},\mu^2) \; 
\sigma_{q \overline q}^{\rm res}(\tau S, m^2) \, , 
\label{hadres}
\end{eqnarray}
where $\sigma_{q \overline q}^{\rm res}(\tau S, m^2)$ is defined in 
Eq.~(\ref{partres}) and $\tau_0 = (m + \sqrt{m^2 + s_{\rm cut}})^2/S$.  

In Fig. 5 we show numerical results for the $t \overline t$ production
cross section at the Fermilab Tevatron with $\sqrt S=1.8$ TeV.
We use the CTEQ 4D DIS parton densities \cite{CTEQ}.
The NLO exact cross sections, including the factorization scale 
dependence, are shown as functions of the top quark mass.
We also plot the NLO approximate cross section, 
i.e. the one-loop expansion of the resummed cross section,
calculated with $s_{\rm cut}=0$ and $\mu^2=m^2$.  
We note the excellent agreement between the NLO exact and
approximate cross sections.

\EPSFIGURE{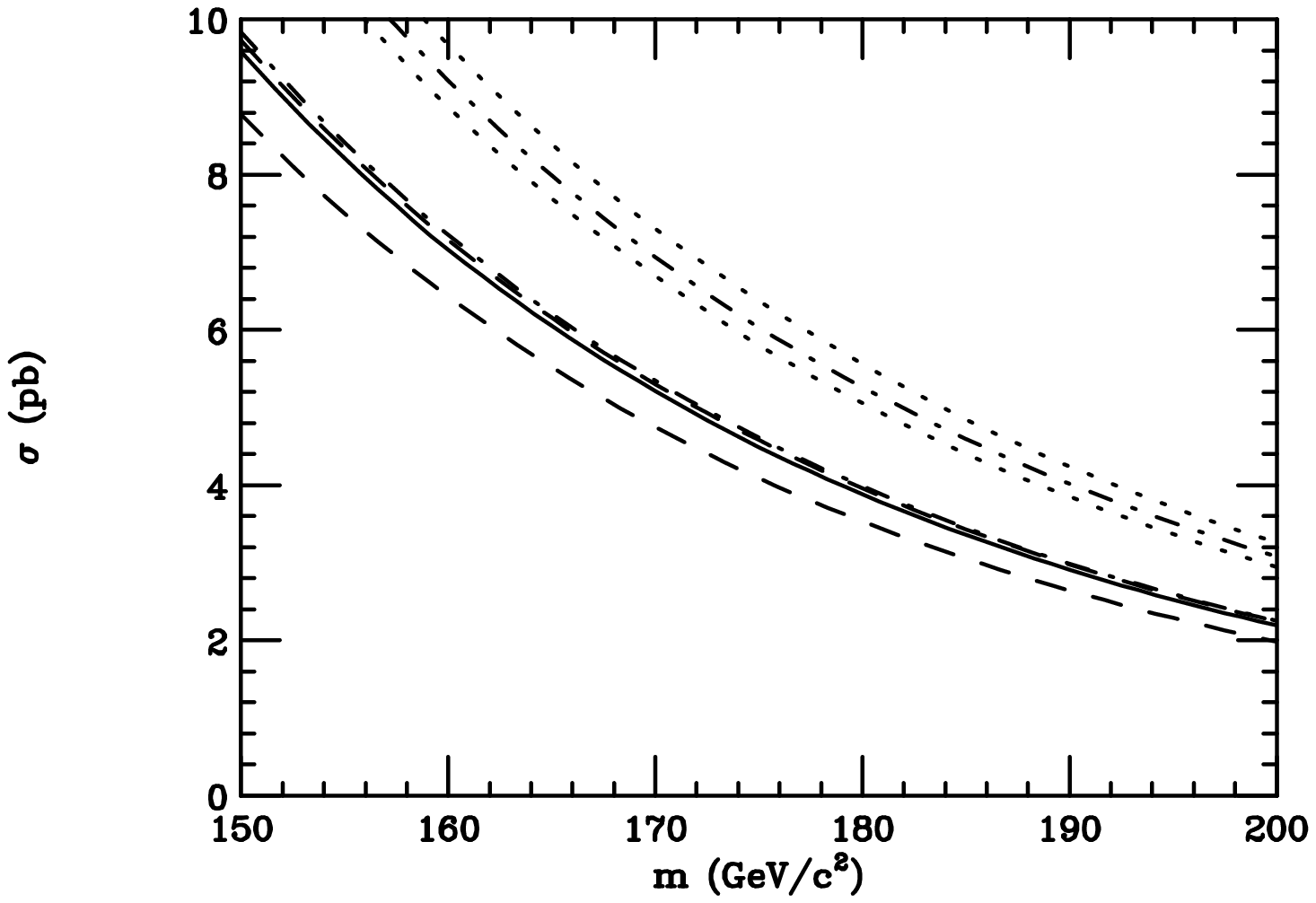,width=5.05in}
{The NLO exact and approximate and the NLL improved 
hadronic $t \overline t$ production cross sections in 
the $q \overline q$ channel and the DIS scheme are
given as functions of top quark mass for $p \overline p$ collisions at the
Tevatron energy, $\sqrt{S} = 1.8$ TeV. The NLO exact cross section is given
for $\mu^2 = m^2$ (solid curve),
$4m^2$ (lower-dashed) and $m^2/4$ (upper-dashed).
The NLO approximate cross section with $s_{\rm cut} = 0$ is shown for
$\mu^2 = m^2$ (lower dot-dashed).
The NLL improved cross section, Eq.~(\ref{improved}), 
is given for $s_{\rm cut} = 35 s_{4, {\rm  min}}$ (upper dot-dashed), 
$30 s_{4, {\rm min}}$ (upper-dotted) and $40s_{4, {\rm min}}$
(lower-dotted).}

The scale dependence of the NLL resummed cross section
is significantly reduced relative to that of the NLO cross section.
In order to match our results to the exact NLO cross section 
we define the NLL improved cross section
\beq
\sigma_{q \overline q, {\rm had}}^{\rm imp} 
= \sigma_{q \overline q, {\rm had}}^{\rm res} -
\sigma_{q \overline q, {\rm had}}^{\rm NLO, approx} 
+ \sigma_{q \overline q, {\rm had}}^{\rm NLO, exact} \, \, ,
\label{improved} 
\eeq
with the same cut applied to the NLO approximate and the
NLL resummed cross sections.

We plot the hadronic improved cross section in Fig. 5
for $\mu^2 = m^2$ along with the variation with $s_{\rm cut}$. 
The variation of the improved cross section with change
of cutoff is small over the range 
$30 s_{4, {\rm min}}< s_{\rm cut} <40 s_{4, {\rm min}}$.
At $m = 175$ GeV/$c^2$ and $\sqrt{S} = 1.8$ TeV, 
the value of the improved cross section for 
$q \overline q \rightarrow t \overline t$ with 
$s_{\rm cut}/(2m^2)=0.04$ is 6.0 pb
compared to a NLO cross section of 4.5 pb at $\mu=m$.
The contribution of the $gg \rightarrow Q {\bar Q}$ channel to top
quark production at the Tevatron is relatively small. 
With its inclusion the total $t {\bar t}$ cross section at the Tevatron
is 7 pb. Gluon fusion is much more important for $b$-quark production
at HERA-B \cite{HERAB}.

\section{Resummation for dijet production}

Here we review the resummation formalism for the
hadronic production of a pair of jets. 
Dijet production entails additional complications
due to the presence of final state jets; otherwise,
the formalism is analogous to that for heavy quark production.

\subsection{Factorized dijet cross section}
We consider dijet production in hadronic processes
\beq
h_a(p_a)+h_b(p_b) \rightarrow J_1(p_1,\delta_1)+J_2(p_2,\delta_2)+X \, ,
\label{dijet}
\eeq
where $\delta_1$ and $\delta_2$ are the cone angles of the jets.
The partonic cross section is infrared safe since
the initial-state collinear singularities 
are factorized into universal parton distribution functions,
and the use of cones removes all the final-state 
collinear singularities. 

We study inclusive dijet cross sections
at fixed rapidity interval
\beq
{\Delta}y= (1/2)\ln [(p_1^+\; p_2^-) /(p_1^-\; p_2^+)] \, ,
\eeq
and total rapidity
\beq
y_{JJ}= (1/2) \ln[(p_1^++p_2^+)/(p_1^-+p_2^-)] \, .
\eeq
To construct the dijet cross sections, we define a large invariant,
$M_{JJ}$, which is held fixed. A natural choice is
the dijet invariant mass,
$M^2_{JJ}=(p_1+p_2)^2$, analogous to $Q^2$ for heavy
quark production, but other choices are possible, such as
the scalar product of the  two jet momenta, $M^2_{JJ}=2p_1 \cdot p_2$.
We note that for large $M_{JJ}$ at fixed $\Delta y$ we have a large 
momentum transfer in the partonic subprocess.
The resummed cross section depends critically 
on the choice for $M_{JJ}$. 
We shall see that the leading behavior of the resummed 
cross section is the same as for heavy quark production
when $M_{JJ}$ is the dijet invariant mass, while it is
different for the other choice. 

The dijet cross section can be written in factorized form as   
\beqa
&&\frac{d\sigma_{h_ah_b{\rightarrow}J_1J_2}}
{dM^2_{JJ}\, dy_{JJ} \, d{\Delta}y}= 
\sum_{f_a,f_b=q,\overline{q},g} 
\int \frac{dx_a}{x_a} \, \frac{dx_b}{x_b}
\nonumber \\ &&  \times \, 
\phi_{f_a/h_a}(x_a,\mu^2) \; \phi_{f_b/h_b}(x_b,\mu^2) 
\nonumber \\ &&  \times \, 
H_{f_af_b}\left(\frac{M_{JJ}^2}{x_ax_bS},y,\Delta y,\frac{M_{JJ}}{\mu}, 
\alpha_s(\mu^2),\delta_1,\delta_2\right) ,
\nonumber \\
\label{dijetfact}
\eeqa
where $H$ is the hard scattering and the $\phi$'s are parton
distribution functions.

As for heavy quark production, we may replace the incoming hadrons by
partons in the above equation,
and integrate over the total rapidity of the jet pair. 
Then, we can write the infrared regularized partonic cross section, 
which factorizes as the hadronic cross section, as
\beqa
&& \hspace{-3mm} \frac{d\sigma_{f_af_b{\rightarrow}J_1J_2}}{dM^2_{JJ} \; 
d{\Delta}y}=\int_{\tau}^1 dz
\int \frac{dx_a}{x_a} \, \frac{dx_b}{x_b}\; 
\nonumber \\ && \hspace{-3mm} \times \,
\phi_{f_a/f_a}(x_a,\mu^2,\epsilon) \,
\phi_{f_b/f_b}(x_b,\mu^2,\epsilon) \, 
\delta\left(z-\frac{M^2_{JJ}}{s}\right) 
\nonumber \\ &&  \hspace{-3mm} \times 
\sum_{\a} \, \hat{\sigma}_{\a}
\left(1-z,\frac{M_{JJ}}{\mu},{\Delta}y,\alpha_s(\mu^2),
\delta_1,\delta_2\right) \, ,
\label{sigpart}
\eeqa
where the argument $\epsilon$ represents the universal 
collinear singularities, 
and we have introduced a simplified hard-scattering function, 
${\hat \sigma}_{\a}$, where ${\a}$ denotes the partonic subprocesses.
The threshold region is given in terms of the variable $z$,
\beq
z\equiv \frac{M^2_{JJ}}{s}\, ,
\eeq
where, as before, $s=x_ax_bS$ with $S=(p_a+p_b)^2$.  
Partonic threshold is at $z_{\rm max}=1$ 
while the lower limit of $z$ is
$z_{\rm min}\equiv\tau=M^2_{JJ}/S$.

By taking a Mellin transform of the rapidity-integrated partonic 
cross section, Eq. (\ref{sigpart}), we reduce  
the convolution to a simple product of moments, 
\beqa 
&& \hspace{-5mm}
\int_0^1 d\tau\; \tau^{N-1}\; 
\frac{d\sigma_{f_af_b{\rightarrow}J_1J_2}}{dM^2_{JJ} \; d{\Delta}y} 
\nonumber \\ && 
= \sum_{\a}{\tilde \phi}_{f_a/f_a}(N,\mu^2,\epsilon) \, 
{\tilde \phi}_{f_b/f_b}(N,\mu^2,\epsilon) 
\nonumber \\ && \quad \times \,
{{\tilde{\sigma}}}_{\a}(N,M_{JJ}/\mu,\alpha_s(\mu^2),\delta_1,
\delta_2)\, ,
\label{moment}
\eeqa
with $\tilde{\sigma}_{\a}(N)$ 
and $\tilde{\phi}(N)$ defined as in Eq.~(\ref{momphisigma}).  
We then factorize the initial-state collinear divergences into the 
light-cone  distribution functions $\phi_{f/f}$, expanded to the
same order in $\alpha_s$ as the partonic cross section, and thus obtain
the perturbative expansion for the infrared-safe hard
scattering function, ${\tilde {\sigma}}_{\a}$. 

\EPSFIGURE{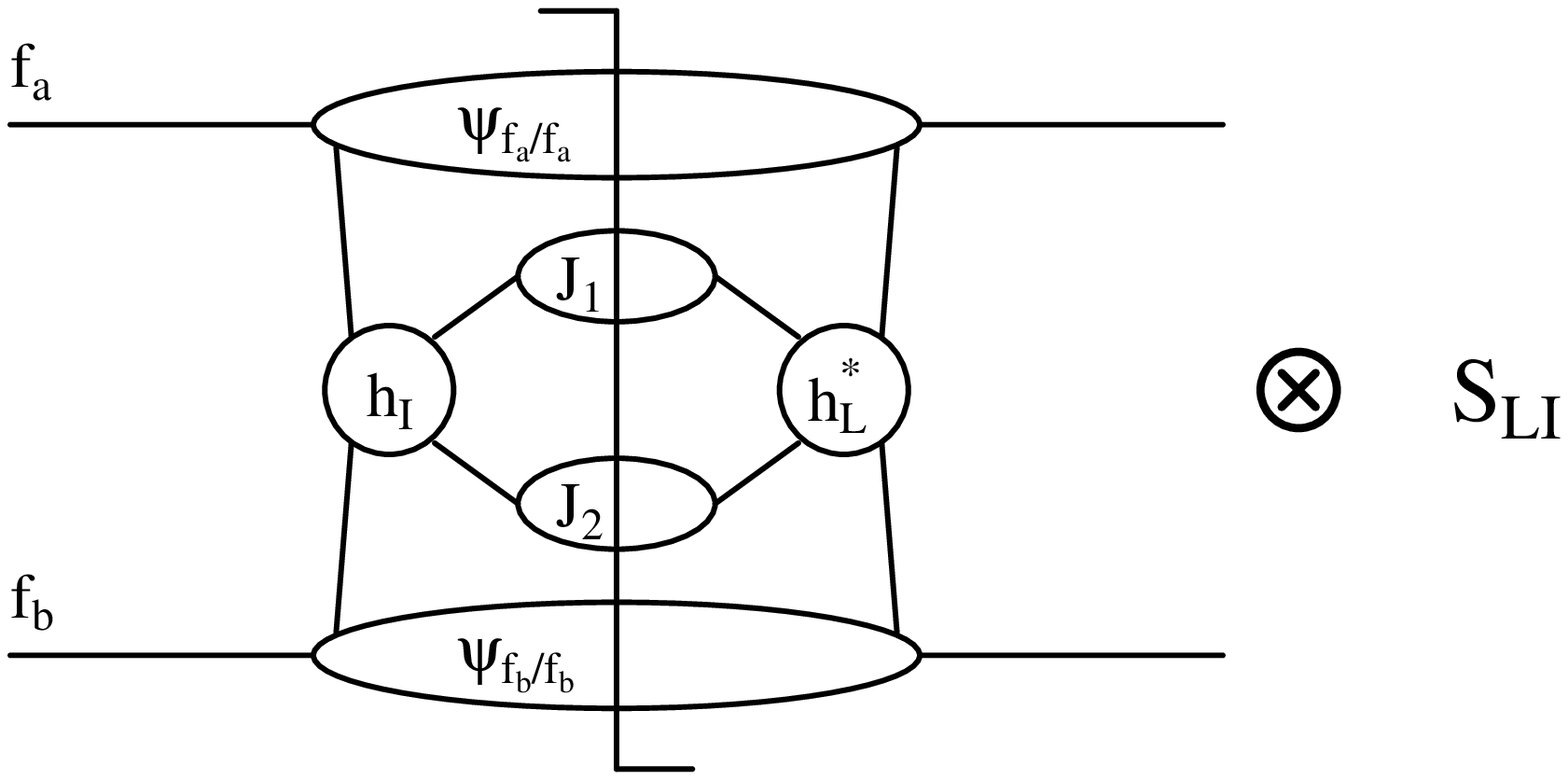,width=5.05in}
{Refactorization for dijet production. The soft gluon function 
is as in Fig. 2.}

We can refactorize the cross section, 
as shown in Fig.~6, into hard components $H_{IL}$,
which describe the truly short-distance hard-scattering,
center-of-mass distributions $\psi$, associated with gluons
collinear to the incoming partons, a soft gluon function
$S_{LI}$, associated with non-collinear soft gluons, 
and jet functions $J_i$, associated with gluons collinear 
to the outgoing jets.
As for heavy quarks, here $I$ and $L$ denote color indices that describe the 
color structure of the hard scattering. 
The refactorized cross section may then be written as
\beqa
&& \int_0^1 d\tau\; \tau^{N-1}\;
\frac{d\sigma_{f_af_b{\rightarrow}J_1J_2}}{dM^2_{JJ} \; d{\Delta}y} 
\nonumber\\ && 
=\sum_{\a}\sum_{IL} 
H_{IL}^{(\a)}\left({M_{JJ}\over\mu},\Delta y \right) \;
{\tilde S}_{LI}^{(\a)} \left({M_{JJ}\over N \mu}, \Delta y \right)
\nonumber\\ && \quad\times\; 
{\tilde\psi}_{f_a/f_a}\left(N,{M_{JJ}\over \mu}, \epsilon \right) \;
{\tilde\psi}_{f_b/f_b}\left(N,{M_{JJ}\over \mu}, \epsilon \right)
\nonumber \\ && \quad \times \; 
{\tilde J}_{(f_1)}\left(N,{M_{JJ}\over \mu }, \delta_1\right)\; 
{\tilde J}_{(f_2)}\left(N,{M_{JJ}\over \mu }, \delta_2\right).
\nonumber \\ .  
\label{refact}
\eeqa
This refactorization is similar to the heavy quark 
case except that we now include in addition  
outgoing jet functions in order to absorb
the final state collinear singularities. 

\subsection{Resummed dijet cross section}

By comparing Eqs.\ (\ref{moment}) and (\ref{refact}), 
we can write the refactorized expression for the Mellin transform of 
the hard-scattering function ${\tilde {\sigma}}_{\a}$ as
\beqa
\tilde{{\sigma}}_{\a}(N)&=&
\left[\frac{ {\tilde{\psi}}_{f_a/f_a}(N,M_{JJ}/\mu)
{\tilde{\psi}}_{f_b/f_b}(N,M_{JJ}/\mu )}
{{\tilde{\phi}}_{f_a/f_a}(N,\mu^2 ) \, 
{\tilde{\phi}}_{f_b/f_b}(N,\mu^2)}
 \right]\nonumber \\ && \hspace{-15mm} \times \,
\sum_{IL}H_{IL}^{(\a)}\left({M_{JJ}\over\mu},\Delta y \right)\; 
{\tilde S}_{LI}^{(\a)} \left( {M_{JJ}\over N \mu},\Delta y \right)
\nonumber \\ && \hspace{-15mm} \times \,
{\tilde J}_{(f_1)}\left(N,{M_{JJ}\over \mu}, \delta_1\right)\; 
{\tilde J}_{(f_2)}\left(N,{M_{JJ}\over \mu}, \delta_2\right).
\label{sigfactjet}
\eeqa

The results for the resummation of the universal ratio $\psi/\phi$ 
and the soft gluon function were given in the context of heavy quark 
production in Section 2. Explicit expressions for the soft anomalous
dimension matrices for dijet production will be given in the next section.
Thus, the only thing remaining to write down the resummed dijet 
cross section are expressions for the resummation of the final state jets. 

The moments of the final-state jet with
$M_{JJ}^2=(p_1+p_2)^2$ are given by~\cite{KOS1} 
\beq
\tilde{J}_{(f_i)}\left(N,{M_{JJ}\over \mu}, \delta_i\right)
=\exp \left[E'_{(f_i)}(N,M_{JJ})\right]\, ,
\label{finaljet}
\eeq
with
\beq
E'_{(f_i)}(N,M_{JJ})=
\int_\mu^{M_{JJ}/N} {d\mu' \over \mu'}\; \; 
C'_{(f_i)}(\alpha_s(\mu'{}^2))\, ,
\label{Epr2exp}
\eeq
where the first term in the series for $C'_{(f_i)}(\alpha_s)$ 
may be read off from a one-loop calculation. 
The leading logarithmic behavior of the 
cross section in this case is not affected by the final state jets, 
so we always have an enhancement of the cross section at leading logarithm, 
as is the case for Drell-Yan and heavy quark cross sections.  

The moments of the final-state jet with
$M_{JJ}^2=2p_1\cdot p_2$ are given by Eq.~(\ref{finaljet})
with~\cite{KOS1}
\beqa
\hspace{-5mm}
E'_{(f)}\left(N,M_{JJ}\right)
&=&
\int^1_0 dz \frac{z^{N-1}-1}{1-z}\; 
\nonumber\\ && \hspace{-25mm} \times 
\left \{\int^{(1-z)}_{(1-z)^2} \frac{d\lambda}{\lambda} 
A^{(f)}\left[\alpha_s(\lambda M_{JJ}^2)\right] \right.
\nonumber\\ &&  \hspace{-25mm} \quad \quad \left.
{}+B'_{(f)}\left[\alpha_s((1-z) M_{JJ}^2) \right] \right\}\, , 
\label{Eprexp}
\eeqa
where the function $A^{(f)}$ is the same as in Eq.\ (\ref{Aexp})
and the lowest-order term in $B'_{(f)}$
may be read off from the one-loop jet function.  
The results include a gauge dependence, which cancels
against a corresponding dependence in the soft anomalous
dimension matrix.
The leading logarithms for final-state jets with 
$M_{JJ}^2=2p_1\cdot p_2$
are negative and give a suppression to the cross section, in contrast
to the initial-state leading-log contributions.

Using Eqs.~(\ref{sigfactjet}), (\ref{psiphimu}), and (\ref{finaljet}),  
together with the evolution of the soft function,
we can write the resummed dijet cross section in moment space as
\beqa
&& \hspace{-3mm}
\tilde{{\sigma}}_{\a}(N) = R_{(f)}^2\; 
\exp \left \{ \sum_{i=a,b} \left[ E^{(f_i)}(N,M_{JJ}) \right. \right. 
\nonumber\\ && \hspace{-6mm}\left. \left.
-2\int_\mu^{M_{JJ}}{d\mu'\over\mu'}
\left[\gamma_{f_i}(\alpha_s(\mu'{}^2))-\gamma_{f_if_i}(N,\alpha_s(\mu'{}^2)) 
\right]\right]\right\}
\nonumber \\ &&
\times\; \exp \left \{\sum_{j=1,2}E'_{(f_j)}(N,M_{JJ}) \right\}
\nonumber\\ && \times\; {\rm Tr} \left\{ 
H^{(\a)}\left({M_{JJ}\over\mu},\Delta y,\alpha_s(\mu^2)\right) \right.
\nonumber\\ && \times\;
\bar{P} \exp \left[\int_\mu^{M_{JJ}/N} {d\mu' \over \mu'}\; 
\Gamma_S^{(\a)}{}^\dagger\left(\alpha_s(\mu'^2)\right)\right]
\nonumber\\ && \times\;
{\tilde S}^{(\a)} \left(1,\Delta y,\alpha_s\left(M_{JJ}^2/N^2\right) \right)
\nonumber\\ && \left. \times\; 
P \exp \left[\int_\mu^{M_{JJ}/N} {d\mu' \over \mu'} \;
\Gamma_S^{(\a)}\left(\alpha_s(\mu'^2)\right)\right] \right\}.
\eeqa
This expression is analogous to Eq.~(\ref{resHQ}) 
for heavy quark production except for the
addition of the exponents for the final-state jets. 

We give explicit expressions for the soft anomalous
dimension matrices, $\Gamma_S$, 
for the partonic subprocesses in dijet
production in the next section.

\section{Soft anomalous dimension matrices for 
dijet production}

We consider partonic subprocesses 
\beq
f_a\left(p_a, r_a \right) + f_b\left(p_b, r_b \right) \rightarrow 
f_1\left(p_1, r_1 \right) + f_2\left(p_2, r_2 \right) ,
\eeq
where the $p_i$'s and $r_i$'s denote momenta and colors of the partons
in the process. 
To facilitate the presentation of the results for $\Gamma_S$ we
use the notation
\beq
\T\equiv \ln\left(\frac{-t}{s}\right)+\pi i, \, 
\U\equiv \ln\left(\frac{-u}{s}\right)+\pi i,
\label{eq:new2form}
\eeq
where
\beq
s=\left(p_a+p_b \right)^2, \, t=\left(p_a-p_1 \right)^2, \,  
u=\left(p_a-p_2 \right)^2,
\label{Mandlst}
\eeq
are the usual Mandelstam invariants. 
We note that for the definition of the color basis we can choose any 
physical channel $s$, $t$, or $u$. 
We will use $t$-channel bases for the partonic processes in dijet
production except for the processes $q {\bar q}\rightarrow gg$ 
and $gg \rightarrow q{\bar q}$  
which are better described in terms of $s$-channel color structures.

Since the full cross section is gauge independent,
the gauge dependence in the product of the hard and soft functions,
$H^{(\a)} \, S^{(\a)}$, 
must cancel the gauge dependence of the incoming jets, $\psi$, 
and outgoing jets, $J_{(f_i)}$.
Since the jets are incoherent relative to the hard and soft functions, 
the gauge dependence of the anomalous dimension matrices $\Gamma_S^{(\a)}$ 
must be proportional to the identity matrix.
Then we can rewrite the anomalous dimension matrices as
\beqa
&& \hspace{-10mm}
(\Gamma^{(\a)}_S)_{KL}=(\Gamma^{(\a)}_{S'})_{KL}
+ \delta_{KL}  \frac{\alpha_s}{\pi} 
\nonumber \\ && \hspace{-10mm} \times 
\sum_{i=a,b,1,2}C_{(f_i)} \, 
\frac{1}{2} \,  (-\ln\nu_i-\ln 2+1-\pi i) \, ,
\label{gammagaug}
\eeqa
with $C_{f_i}=C_F\ (C_A)$ for a quark (gluon), and with
$\nu_i \equiv (v_i \cdot n)^2/|n|^2$.
Here the dimensionless and lightlike velocity vectors $v_i^{\mu}$ are
defined by $p_i^{\mu}=M_{JJ}v_i^{\mu}/{\sqrt{2}}$ 
and satisfy $v_i^2=0$.

In the following subsections we will present
the explicit expressions for $\Gamma^{(\a)}_{S'}$
for all the partonic subprocesses in dijet production,
by evaluating one-loop diagrams as in Fig.~3. The full anomalous
dimension matrices can be retrieved from Eq.~(\ref{gammagaug}).

\subsection{Soft anomalous dimension for $q \bar{q}\rightarrow q \bar{q}$}

First, we present the soft 
anomalous dimension matrix for the process
\beq
q\left(p_a, r_a \right)+\bar{q}\left(p_b, r_b \right) \rightarrow
q\left(p_1, r_1 \right)+\bar{q}\left(p_2, r_2 \right) \, ,
\eeq
in the $t$-channel singlet-octet color basis
\beqa
c_1&=&\delta_{r_a r_1}\delta_{r_b r_2} \, , \nonumber\\
c_2&=&(T_F^c)_{r_1 r_a}(T_F^c)_{r_b r_2}.
\eeqa
We find~\cite{KOS2,BottsSt}
\beq
\Gamma_{S'}=\frac{\alpha_s}{\pi}\left[
                \begin{array}{cc}
                 2{C_F}\T  &   -\frac{C_F}{N_c} \U  \vspace{2mm} \\
                -2\U    &-\frac{1}{N_c}(\T-2\U)
                \end{array} \right]\, .
\eeq

We note that the dependence on $\T$ is diagonal in 
this $t$-channel color basis and in the forward region of 
the partonic scattering, $\T\rightarrow -\infty$, 
where $\Gamma_{S}$ becomes diagonal,
color singlet exchange is exponentially enhanced relatively to color octet.

\subsection{Soft anomalous dimension for $q q\rightarrow q q$
and ${\bar q}{\bar q}\rightarrow {\bar q}{\bar q}$}
Next, we consider the process
\beq 
q\left(p_a, r_a \right)+q\left(p_b, r_b \right) \rightarrow
q\left(p_1, r_1 \right)+q\left(p_2, r_2 \right) \, ,
\eeq
in the $t$-channel singlet-octet color basis
\beqa
c_1&=&(T_F^c)_{r_1 r_a}(T_F^c)_{r_2 r_b} \, ,
\nonumber\\
c_2&=&\delta_{r_a r_1} \delta_{r_b r_2}.
\eeqa
The anomalous dimension matrix is~\cite{KOS2,BottsSt}
\beq
\Gamma_{S'}=\frac{\alpha_s}{\pi}\left[
                \begin{array}{cc}
                -\frac{1}{N_c}(\T+\U)+2C_F \U  &  2\U \vspace{2mm} \\
                 \frac{C_F}{N_c} \U    & 2{C_F}\T
                \end{array} \right]
\eeq
which also applies to the process
\beq 
{\bar q}\left(p_1, r_1 \right)+{\bar q}\left(p_2, r_2 \right) \rightarrow
{\bar q}\left(p_a, r_a \right)+{\bar q}\left(p_b, r_b \right) \, .
\eeq

Again, we note that the dependence on $\T$ is diagonal in this
$t$-channel color basis, and the color singlet dominates  
in the forward region of the partonic scattering.

\subsection{Soft anomalous dimension for $q \bar{q}\rightarrow g g$ and
$g g \rightarrow q \bar{q}$}

Here, we present the soft anomalous dimension matrix for 
the process
\beq
q\left(p_a, r_a \right)+\bar{q}\left(p_b, r_b \right) \rightarrow
g\left(p_1, r_1 \right)+g\left(p_2, r_2 \right) \, ,
\eeq
in the $s$-channel color basis
\beqa
c_1&=&\delta_{r_a r_b}\delta_{r_1 r_2} \, ,
\nonumber \\ 
c_2&=&d^{r_1 r_2 c}{\left( T_F^c \right)}_{r_b r_a} \, ,
\nonumber \\
c_3&=&if^{r_1 r_2 c}{\left( T_F^c \right)}_{r_b r_a} \, .
\eeqa
We find~\cite{KOS2} 
\beqa
&&\Gamma_{S'}=\frac{\alpha_s}{\pi}
\\ && \nonumber \times
\left[
                \begin{array}{ccc}
                 0  &   0  & \U-\T  \vspace{2mm} \\ 
                 0  &   \frac{C_A}{2}\left(\T+\U \right)    & \frac{C_A}{2}
\left(\U-\T\right) \vspace{2mm} \\ 
                 2\left(\U-\T \right)  & \frac{N_c^2-4}{2N_c}\left(\U-\T 
\right)  & \frac{C_A}{2}\left(\T+\U \right)
                \end{array} \right].
\label{Gammaqqgg}
\eeqa
The same anomalous dimension describes also the time-reversed process
\cite{Thesis,KS}
\beq
g\left(p_1, r_1 \right)+g\left(p_2, r_2 \right) \rightarrow
\bar{q}\left(p_a, r_a \right)+q\left(p_b, r_b \right) .
\eeq

\subsection{Soft anomalous dimension for $qg \rightarrow qg$ and
$\bar{q} g \rightarrow \bar{q} g$}

Next, we consider the ``Compton'' process
\beq
q\left(p_a, r_a \right)+g\left(p_b, r_b \right) \rightarrow
q\left(p_1, r_1 \right)+g\left(p_2, r_2 \right) \, ,
\eeq
in the $t$-channel color basis
\beqa
c_1&=&\delta_{r_a r_1}\delta_{r_b r_2} \, ,  
\nonumber \\
c_2&=&d^{r_b r_2 c}{\left( T_F^c \right)}_{r_1 r_a} \, ,
\nonumber \\
c_3&=&if^{r_b r_2 c}{\left( T_F^c \right)}_{r_1 r_a} \, .
\label{eq:basqgqg}
\eeqa
The soft anomalous dimension matrix is  \cite{KOS2}
\beqa
&&\Gamma_{S'}=\frac{\alpha_s}{\pi}
\\ && \nonumber \times
\left[
                \begin{array}{ccc}
                 \left( C_F+C_A \right) \T  &   0  & \U  \vspace{2mm} \\ 
                 0  &   C_F \T+ \frac{C_A}{2} \U     & \frac{C_A}{2} \U  
\vspace{2mm} \\
                 2\U  & \frac{N_c^2-4}{2N_c}\U  &  C_F \T+ \frac{C_A}{2}\U
                \end{array} \right] 
\label{Gammaqgqg}
\eeqa
which also applies to the process
\beq
\bar{q}\left(p_1, r_1 \right)+g\left(p_2, r_2 \right) \rightarrow
\bar{q}\left( p_a, r_a \right)+g\left(p_b, r_b \right) \, .
\eeq

We note that $T$ appears only in the diagonal of the anomalous
dimension matrix and that in the forward region 
($T \rightarrow -\infty$) the color singlet dominates.
 
\subsection{Soft anomalous dimension for $gg \rightarrow gg$}

Finally, we consider the much more complicated process 
\beq
g\left(p_a, r_a \right)+g\left(p_b, r_b \right) \rightarrow
g\left(p_1, r_1 \right)+g\left(p_2, r_2 \right) \, .
\eeq
A complete color basis for this process 
is given by the eight color structures~\cite{KOS2}
\beqa
c_1&=&\frac{i}{4}\left[f^{r_a r_b l}
d^{r_1 r_2 l} - d^{r_a r_b l}f^{r_1 r_2 l}\right] \, ,
\nonumber \\
c_2&=&\frac{i}{4}\left[f^{r_a r_b l} 
d^{r_1 r_2 l} + d^{r_a r_b l}f^{r_1 r_2 l}\right] \, ,
\nonumber \\ 
c_3&=&\frac{i}{4}\left[f^{r_a r_1 l}
d^{r_b r_2 l}+d^{r_a r_1 l}f^{r_b r_2 l}\right] \, , 
\nonumber \\
c_4&=&P_1(r_a,r_b;r_1,r_2)=\frac{1}{8}\delta_{r_a r_1} 
\delta_{r_b r_2} \, ,
\nonumber \\
c_5&=&P_{8_S}(r_a,r_b;r_1,r_2)
=\frac{3}{5} d^{r_ar_1c} d^{r_br_2c} \, ,
\nonumber \\
c_6&=&P_{8_A}(r_a,r_b;r_1,r_2)=\frac{1}{3} f^{r_ar_1c} f^{r_br_2c} \, ,
\nonumber \\
c_7&=&P_{10+{\overline {10}}}(r_a,r_b;r_1,r_2)=    
\frac{1}{2}(\delta_{r_a r_b} \delta_{r_1 r_2} 
\nonumber \\ && \quad
{}-\delta_{r_a r_2} \delta_{r_b r_1})
-\frac{1}{3} f^{r_ar_1c} f^{r_br_2c} \, ,
\nonumber \\
c_8&=&P_{27}(r_a,r_b;r_1,r_2)=\frac{1}{2}(\delta_{r_a r_b} \delta_{r_1 r_2}
\nonumber \\ && \hspace{-3mm}
{}+\delta_{r_a r_2} \delta_{r_b r_1})
-\frac{1}{8}\delta_{r_a r_1} \delta_{r_b r_2}
-\frac{3}{5} d^{r_ar_1c} d^{r_br_2c} \, ,
\nonumber \\
\label{8x8basis}
\eeqa
where the $P$'s are $t$-channel projectors of irreducible
representations of $SU(3)$ \cite{Bart}, and we use explicitly $N_c=3$.

The soft anomalous dimension matrix in this basis is~\cite{KOS2}
\beq
\Gamma_{S'}=\left[\begin{array}{cc}
            \Gamma_{3 \times 3} & 0_{3 \times 5} \\
              0_{5 \times 3}      & \Gamma_{5 \times 5}
\end{array} \right] \, ,
\label{gammagggg}
\eeq
with
\beq
\blocA=\frac{\alpha_s}{\pi} \left[
                \begin{array}{ccc}
                  3T  &   0  & 0  \\
                  0  &  3U & 0    \\
                  0  &  0  &  3\left(\T+\U \right)
                   \end{array} \right]
\eeq
and
\beqa
&&\Gamma_{5 \times 5}=\frac{\alpha_s}{\pi}
\\ && \nonumber \hspace{-3mm} \times
\left[\begin{array}{ccccc}
6\T & 0 & -6\U & 0 & 0 \vspace{2mm} \\ 
0  & 3\T+\frac{3\U}{2} & -\frac{3\U}{2} & -3\U & 0 \vspace{2mm} \\ 
-\frac{3\U}{4} & -\frac{3\U}{2} &3\T+\frac{3\U}{2} & 0 & -\frac{9\U}{4} 
\vspace{2mm} \\
0 & -\frac{6\U}{5} & 0 & 3\U & -\frac{9\U}{5} \vspace{2mm} \\
0 & 0 &-\frac{2\U}{3} &-\frac{4\U}{3} & -2\T+4\U
\end{array} \right] \, .
\eeqa
We note that the dependence on $\T$ is diagonal and that
color singlet exchange dominates in the forward region
of the partonic scattering, $\T \rightarrow -\infty$, 
where $\Gamma_{S}$ becomes diagonal.
This has been the case for all the processes
we analyzed in $t$-channel bases; suppression increases with the dimension
of the exchanged color representation.

The calculation of the eigenvalues and eigenvectors of $\Gamma_{S'}$
for the $gg \rightarrow gg$ process is more difficult than for the
other partonic processes since we are now dealing with a $8 \times 8$
matrix; however, explicit results have been obtained \cite{KOS2}.
The eigenvalues of the anomalous dimension matrix, 
Eq.~(\ref{gammagggg}), are
\beqa
\lambda_1&=&\lambda_4=3 \, \frac{\alpha_s}{\pi} \, \T \, ,
\nonumber \\ 
\lambda_2&=&\lambda_5=3 \, \frac{\alpha_s}{\pi} \, \U \, ,
\nonumber \\
\lambda_3&=&\lambda_6=3 \, \frac{\alpha_s}{\pi} \, (\T+\U) \, , 
\nonumber \\
\lambda_{7,8}&=&2 \, \frac{\alpha_s}{\pi} \left[\T+\U \right.
\nonumber \\ && \left. \,
\mp 2\sqrt{\T^2-\T\U+\U^2} \right] \, .
\eeqa
The eigenvectors have the general form
\beqa
e_i&=&\left[\begin{array}{c}
      e_i^{(3)} \\ 
       0^{(5)} 
\end{array}\right], \; i=1,2,3 \, , 
\nonumber \\
e_i&=&\left[\begin{array}{c}
     0^{(3)} \\ 
     e_i^{(5)} 
\end{array}\right], \; i=4 \ldots 8 \, ,
\eeqa
where the superscripts refer to the dimension.
The three-dimensional vectors $e_i^{(3)}$ are given by
\beq
e_i^{(3)}=\left[\begin{array}{c}
     \delta_{i1} \\ 
     \delta_{i2} \\  
        \delta_{i3} 
\end{array}\right], \; i=1,2,3 \, .
\label{ei3}
\eeq
The five-dimensional vectors
$e_4^{(5)}$,  $e_5^{(5)}$ and 
$e_6^{(5)}$ are given by~\cite{KOS2}
\beqa
e_4^{(5)}&=&\left[\begin{array}{c}
      -15 \vspace{2mm} \\ 
     6-\frac{15}{2}\frac{\T}{\U} \vspace{2mm} \\
      -\frac{15}{2}\frac{\T}{\U} \vspace{2mm} \\ 
        3 \vspace{2mm} \\ 
        1 \end{array} \right], \; \;  
e_5^{(5)}=\left[\begin{array}{c}
      0   \vspace{2mm} \\
      -\frac{3}{2} \vspace{2mm} \\
       0  \vspace{2mm} \\
      \frac{3}{4}-\frac{3}{2}\frac{\T}{\U} \vspace{2mm}  \\
       1 \end{array} \right], 
\nonumber \\
e_6^{(5)}&=&\left[\begin{array}{c}
      -15  \vspace{2mm} \\
     -\frac{3}{2}+\frac{15}{2}\frac{\T}{\U} \vspace{2mm}  \\ 
      \frac{15}{2}-\frac{15}{2}\frac{\T}{\U} \vspace{2mm} \\ 
        -3 \vspace{2mm} \\ 
        1 \end{array} \right] \, .
\eeqa
The expressions for $e_7^{(5)}$ and $e_8^{(5)}$ are long but can be given 
succinctly by~\cite{KOS2} 
\beqa
e_i^{(5)}&=&  \left[\begin{array}{c}
        b_1(\lambda_i') \vspace{2mm} \\
        b_2(\lambda_i') \vspace{2mm} \\ 
        b_3(\lambda_i') \vspace{2mm} \\
        b_4(\lambda_i') \vspace{2mm} \\
        1 
\end{array} \right], \; i=7,8 \, ,   
\eeqa
where 
\beq
\lambda_i'=\frac{\pi}{\alpha_s} \lambda_i \, ,
\eeq 
and where the $b_i$'s are given by
\beqa
b_1(\lambda_i')&=&\frac{3}{\U^2 K'} [80 \T^4+103 \U^4-280 \U \T^3 
\nonumber \\ && \quad \quad \; \; 
{}-300 \T \U^3 +404 \T^2 \U^2
\nonumber \\ && \hspace{-12mm}
{}+(40 \T^3-16 \U^3 -60 \T^2 \U +52 \T\U^2)\lambda_i'] \, ,
\nonumber \\
b_2(\lambda_i')&=&\frac{3}{2K'}[20 \T^2-50 \U\T +44 \U^2
\nonumber \\ && \quad \quad
{}+(10 \T-5 \U)\lambda_i'] \, ,
\nonumber \\
b_3(\lambda_i')&=&-\frac{3}{2 \U K'}[40\T^3-64\U^3-120 \T^2 \U
\nonumber \\ && \hspace{-12mm}
{}+130 \T \U^2+(20 \T^2+13 \U^2-20 \T\U) \lambda_i'] \, ,
\nonumber \\
b_4(\lambda_i')&=&\frac{3\U}{K'}(2\T+5\U-2\lambda_i') \, ,
\eeqa
with
\beq
K'=20 \T^2-20 \U\T+21 \U^2 \, .
\eeq

\section{Conclusion}

We have discussed soft gluon resummation at next-to-leading 
logarithmic accuracy for heavy quark and dijet production
in hadronic collisions. 
We have constructed the resummed cross sections in terms
of exponentials of soft anomalous dimension matrices which 
describe the factorization of noncollinear soft gluons from 
the hard scattering.
We have presented explicit results for the soft anomalous 
dimension matrices for the relevant partonic subprocesses. 
Numerical results have been presented for 
top quark production at the Fermilab Tevatron. 
Similar resummations have been recently applied to
direct photon \cite{NK,LOS,NKJO} and $W$ + jet \cite{NK,NKVD}
production, as well as to calculations
of transverse momentum and rapidity distributions \cite{KLMV,pty}
for heavy quark production.

\end{document}